# Extraterrestrial Photochemistry: Principles and Applications[a]

Christopher R. Arumainayagam,[1,b] Eric Herbst,[2,3] A.N. Heays,[4]
Ella Mullikin,[1] Megan Farrah,[1] and Michael G. Mavros,[1]

[1] Department of Chemistry, Wellesley College, Wellesley, MA 02481, USA

[2] Department of Chemistry, University of Virginia Charlottesville, VA 22904, USA

[3] Department of Astronomy, University of Virginia, Charlottesville, VA 22904, USA

[4] J. Heyrovský Institute of Physical Chemistry, Academy of Sciences of the Czech Republic

Dolejškova 3, 18223 Prague 8, Czech Republic

## Abstract

Energetic processing of interstellar ice mantles and planetary atmospheres via photochemistry is a critical mechanism in the extraterrestrial synthesis of prebiotic molecules. Photochemistry is defined as chemical processes initiated by photon-induced electronic excitation, not involving ionization. In contrast, photons with energies above the ionization threshold initiate radiation chemistry (radiolysis). Vacuum-ultraviolet (6.2–12.4 eV) light may initiate photochemistry and radiation chemistry because the threshold for producing secondary electrons is lower in the condensed phase than in the gas phase. Approximately half of cosmic-ray induced photons incident on interstellar ices in star-forming regions initiate photochemistry while the rest initiate radiation chemistry. While experimental techniques such as velocity map imaging may be used to extract exquisite details about gas-phase photochemistry, such detailed information cannot be obtained for condensed-phase photochemistry, which involves greater complexity, including the production of excitons, excimers, and exciplexes. Because a primary objective of chemistry is to provide molecular-level mechanistic explanations for macroscopic phenomena, our ultimate goal in this book chapter is to critically evaluate our current understanding of the photochemistry that likely leads to the synthesis of extraterrestrial prebiotic molecules.

**Extraterrestrial Photochemistry: Principles and Applications**

Photochemistry involves reactions of electronically-excited species that are produced by the absorption of non-ionizing photons. In contrast to rotational and vibrational excitations, electronic excitations correspond to excitation energies similar in magnitude to bond dissociation energies and thus facilitate bond cleavage. Photochemistry occurs throughout the universe, including in our galaxy, the Milky Way. Galaxies contain matter in stellar and interstellar environments, the latter consisting partially of interstellar clouds of gas and dust particles replete with molecules in both gaseous and condensed phases. In the colder regions of clouds, dust particles are covered by ice mantles comprised of water and a variety of other molecules, including organic species. Dense interstellar clouds are also the only known location of stellar and planetary formation. As temperature and density increase, stellar genesis progresses through various evolutionary environments, including starless cores, pre-stellar cores, protostellar cores, and hot cores. According to a recent review, "Interstellar ice photochemistry is an efficient pathway to chemical complexity in space. It is a source of prebiotic amino acids and sugars and may be the original source of enantiomeric excess on the nascent Earth."[1] In recent years, laboratory experiments and theoretical calculations have suggested that condensed phase and gas-phase photochemistry play vital roles in the synthesis of prebiotic molecules in a variety of extraterrestrial sources such as (1) the surface and ice mantles of interstellar (sub)micron-size grains,[2-4] (2) gases in circumstellar envelopes associated with the last phase of stellar evolution, and (3) ices and atmospheres of extra-solar and solar planets (e.g., Mars), dwarf planets (e.g., Pluto), moons (e.g., Europa), asteroids, and comets. Extraterrestrial energetic processing of cosmic ices via circularly-polarized UV photons may induce chiral selective chemistry,[5] which may explain the "handedness" of biological molecules, one of the great mysteries surrounding the origin of life. The idea that organic building blocks of life originate in outer space is the fundamental assumption of molecular panspermia, which posits that endogenous delivery of prebiotic molecules kick-started abiogenesis on Earth (Figure 1). Model calculations[6] suggest that during early and late heavy bombardment,[7] intact delivery was possible of HCN, a feedstock molecule for the synthesis of sugars, nucleotides, amino acids, and lipids. Although there is currently no known uninterrupted synthetic sequence to create a protocell from such feedstock molecules delivered to primitive Earth, recent work proposes viable discontinuous mechanisms including (1) water flow chemistry in meteoritic craters[8] and (2) wet-dry cycling of volcanic hot spring pools.[9] The formation of "lyfe"[10] capable of "dissipation, autocatalysis, homeostasis, and learning" likely included photochemistry as a critical step in non-terrestrial environments.

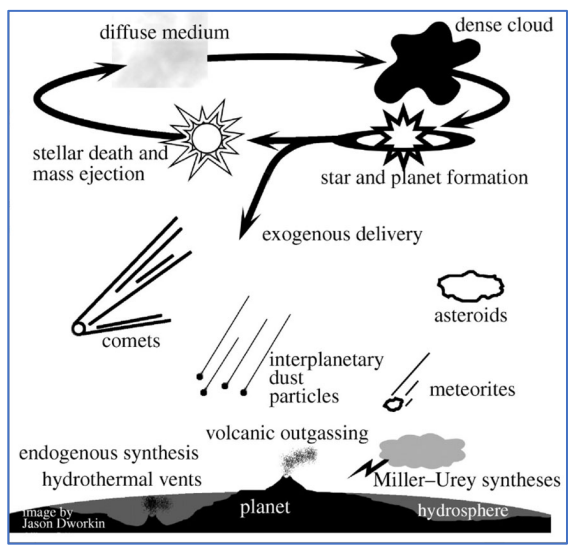

**Figure 1**
Exogenous delivery versus endogenous synthesis.[11]
The publisher for this copyrighted material is Mary
Ann Liebert, Inc. publishers.

Extraterrestrial energetic processing to form prebiotic molecules via photochemistry is the focus of this book chapter, which is subdivided into four sections: (1) introduction: photochemistry in space, (2) photochemistry principles, (3) photochemistry dynamics, and (4) photochemistry in the condensed phase. A strong emphasis is placed on fundamental concepts (e.g., conical intersections, predissociation) and underlying physical processes, some of which may occur within femtoseconds. A strenuous effort is made not to conflate photochemistry with radiation chemistry. For example, while photodissociation by far-UV photons is dominated by electronic excitation (photochemistry), photodissociation by X-ray photons is governed by ionization (radiation chemistry). Therefore, photochemistry is a subset of photon-initiated chemistry. We will examine the difference in chemical effects of low-energy secondary electrons[12]—agents of radiation chemistry—and low-energy (< 10 eV) photons—instigators of photochemistry.

1. Introduction: Photochemistry in Space

**1.1. Photons in space**

Photochemistry in space is initiated by three non-ionizing (typically < 10 eV) photon sources: (1) the interstellar radiation field (ISRF), (2) protostellar and stellar blackbody radiation, and (3) secondary UV

radiation.[c] A conspicuous difference between gas-phase interstellar photochemistry and atmospheric photochemistry arises because of the significant difference in particle collision frequencies in these two environments. The scarcity of collisions in interstellar space allows for the radiative relaxation of electronically-excited photodissociation products (e.g., excited radicals such as •$CH_2OH^*$) before they engage in ongoing chemistry. Within dense planetary atmospheres, in contrast, the extra internal energy of electronically-excited fragments may be available in the next collision.

A short description (e.g., photon flux and energy range) of each extraterrestrial photon source is given below.

**1.1.1. Interstellar radiation field (ISRF)**

The interstellar radiation field (ISRF), a photon field averaged over all types of stars and dust of assorted temperatures, permeates all of space except within dark, dense molecular clouds. The dominant component of the ISRF is the continuous infrared emission from warm dust particles composed mainly of silicates or amorphous carbon. The estimated average UV photon flux in the ISRF between 912–3000 Å is $10^5$ photons/cm$^2$/sec/Å, which gives an integrated flux of ~ $10^8$ photons/cm$^2$/sec between 4.1 and 13.6 eV. There is virtually no UV light above 13.6 eV, the ionization energy of atomic hydrogen (H), because of absorption by neutral H surrounding stellar sources of UV radiation. Star-forming regions are an exception as they are rich in young, massive stars, where the far-UV intensity may be enhanced by orders of magnitude. The ISRF likely does not contribute to the energetic processing of ice mantles, found preferentially in the colder regions of clouds, because of absorption by the surrounding dust cloud. In contrast, the ISRF plays a vital role in the photochemistry occurring in the outer layers of interstellar clouds, protoplanetary disks,[d] and circumstellar envelopes.[e]

**1.1.2. Protostellar and stellar blackbody radiation**

Stellar blackbody UV radiation influences the chemistry of protoplanetary and planetary objects. Most UV radiation from a protostar will be absorbed by nearby dust. However, at infrared (IR) wavelengths, photons can penetrate further and heat dust and ice close to the protostar. As a result, it is unlikely that any ice exists on grains that receive substantial UV radiation. Therefore, protostellar UV radiation will likely not contribute to the energetic processing of grains in hot cores/hot corinos at 100–300 K. The ices will have sublimated at these temperatures into the gas phase, where high abundances of terrestrial-like organic molecules are observed. In contrast, in protoplanetary disks, stellar blackbody radiation from young stars,[f] including UV photons as well as intense bursts of stellar X-rays, lead to photochemistry as well as radiation chemistry. Similarly, blackbody radiation from the outer layers of the Sun, along with the solar wind of high-energy (up to10 keV) particles, is responsible for the photochemistry and radiation chemistry of solar system planetary objects.

---

[c] Protostars are stars in the act of formation before the so-called main sequence of hydrogen burning in their interiors.
[d] Protoplanetary disks, surrounding protostars of young low-mass stars, are formed from the accretion of interstellar dust particles.
[e] Circumstellar envelopes are extended atmospheres of older stars, such as asymptotic giant branch (AGB) stars.

**1.1.3. Secondary UV radiation**

The source of UV light that initiates chemical reactions within cold starless cores and their environs, known as dark clouds, is thought to be local because these clouds are opaque to external UV light from the interstellar radiation field. The UV photons are formed interior to these molecular clouds where cosmic rays, mainly comprised of protons with energies between 10 and 100 MeV, ionize molecular hydrogen to generate secondary electrons, with mean energies around 30 eV.[13] These low-energy secondary electrons and primary cosmic rays excite the Lyman and Werner band systems of molecular hydrogen. The subsequent relaxation of the generated excited states leads to UV emission, whose spectrum is shown in Figure 2. The UV spectrum within dark clouds is, therefore, very similar to that of microwave-discharge hydrogen-flow lamps (MDHLs) used in typical "photochemistry" experiments.[14] However, the flux of secondary UV photons is estimated to be $10^3$ photons cm$^{-2}$ sec$^{-1}$, which is significantly smaller than that of MDHLs used in the laboratory to simulate UV radiation in dark, dense molecular clouds.

---

[f] Young stars are categorized as T Tauri stars, Herbig Ae/Be stars, or massive young stellar objects.

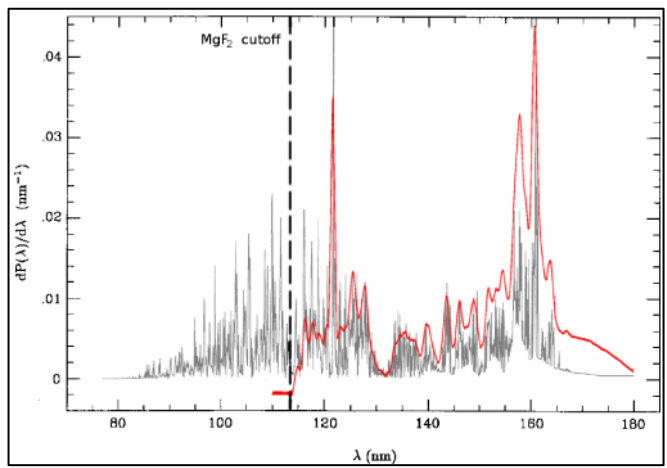

**Figure 2**
Calculated secondary UV radiation spectrum inside dark, dense molecular clouds (black) and spectrum of microwave-discharge hydrogen flow lamp (red).[14] Reproduced with permission © ESO

The secondary UV emission spectrum is dominated by the Lyman-α peak (10.2 eV) and the Lyman band system of molecular hydrogen (6.9–10.9 eV) with peaks at 7.7 and 7.9 eV. *Because only 60% of the flux of secondary UV photons is estimated to be non-ionizing (< 10 eV), this UV light, when interacting with interstellar ices, can produce secondary electrons and cations. Therefore, secondary UV radiation and MDHL lamps typically initiate both photochemistry and radiation chemistry in cosmic ices and their analogs.*

**1.2. Gas-phase extraterrestrial photochemistry**

Photochemistry in the gas-phase occurs in extraterrestrial locations such as protoplanetary disks, planets, lunar atmospheres, and circumstellar envelopes. The resulting radicals may react further with each other and the ambient chemical environment via recombination, disproportionation, and/or recombination/fragmentation. Gas-phase photochemistry is studied using techniques such as flash photolysis, microwave discharge flows, mass spectrometry, synchrotron radiation, plasma dissociation, Laval expansion, crossed molecular beam techniques, and computational chemistry. Velocity map imaging is perhaps the most sensitive tool to study photodissociation dynamics. In this technique, the product atoms undergo resonance-enhanced multi-photon ionization (REMPI), and a two-dimensional detector is used to record the arrival positions of the ions. Analysis of this data yields the product velocity, angular distribution, and the polyatomic co-product's internal energy.[15]

**1.2.1. Gases in protoplanetary disks**

Gaseous UV photochemistry plays a vital role in the chemical processing of protoplanetary disks, which exist around Class II protostars and young stars. Schematically shown in Figure 3, these disks, which are the birthplaces of exo-planets, evolve from protostellar disks associated with Class 0/1 protostars.[16] Recent high-resolution radio observations of the outer layers of some protoplanetary disks reveal concentric gaps, which may indicate the presence of planets.[17] An example of protoplanetary disk chemistry concerns cyanides, the presence of which has been attributed to the interaction of UV light with gases depleted of oxygen, principally hydrogen cyanide (HCN).[18] Enhanced photodissociation of gaseous $^{15}N^{14}N$ compared with $^{14}N_2$ due to self-shielding leads to the incorporation of $^{15}N$ into molecules such as hydrogen cyanide in protoplanetary disks.[19]

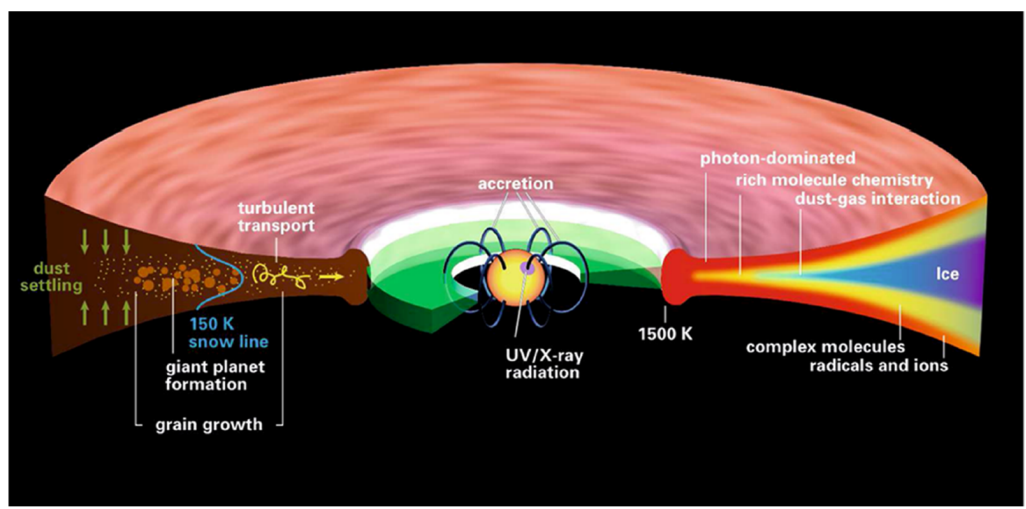

**Figure 3**
Schematic of a protoplanetary disk.[16]
Reprinted with permission.[16] Copyright 2013 American Chemical Society

### 1.2.2. Gases in atmospheres of planets and moons

Gas-phase photochemistry of methane yields prebiotically-significant tholins on Titan, the moon in the solar system with the most substantial atmosphere. The energetic processing of Titan's atmosphere is dominated by solar photons with intensities a hundred-fold less than on Earth.[20] Photodissociation of $N_2$ in Titan's atmosphere leads to the production of nitriles[21] and vinyl cyanide, a potential precursor of cell membranes/vesicle structures.[22] The photochemistry of methane likely occurs in Mars's atmosphere and may explain the conflicting methane measurements from ground-based telescopes, orbiters, and rovers.[23] Host star-driven photochemistry in exoplanetary atmospheres may have implications for the detection of biosignatures on exoplanets.[24]

### 1.2.3. Gases in circumstellar envelopes

Gas-phase photochemistry occurs in circumstellar envelopes (Figure 4) associated with asymptotic giant branch (AGB) stars (both carbon-rich and oxygen-rich), in which helium burning begins in the last phase of stellar evolution of a low- to intermediate-mass (0.8–8 solar mass) star. The external interstellar radiation field, not the internal UV irradiation of the He/H burning shell, photoprocesses the gases ejected by these red giants.[25] Once the central star reaches $3\times10^8$ K on the way to becoming a white dwarf, the intense internal UV radiation induces photochemistry in the resulting planetary nebula[g], facilitating the production of complex molecules that possibly include polycyclic aromatic hydrocarbons[h] (PAHs).[26] This rich chemistry may provide ingredients (dust and gas) for the later synthesis of prebiotic molecules in dark, dense molecular clouds during a new stellar cycle.

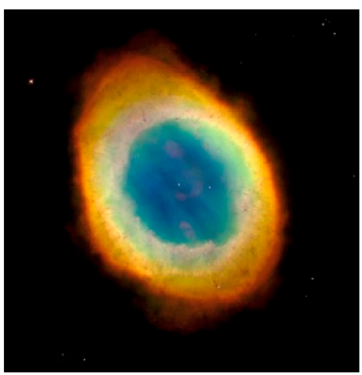

**Figure 4**
Ring Nebula, an example of a planetary nebula.
Image PRC99-01, Space Telescope Science Institute,
Hubble Heritage Team, AURA/STScI/NASA.

**1.3. Condensed-phase extraterrestrial photochemistry**

Both photochemistry and radiation chemistry contribute to the energetic processing of cosmic ices found in interstellar objects, planets, moons, asteroids, and comets.

**1.3.1.   Interstellar condensed-phase photochemistry**

---

[g] When these objects were first discovered in 1784, they were mistakenly associated with planets.
[h] PAHs consist of concatenated aromatic rings.

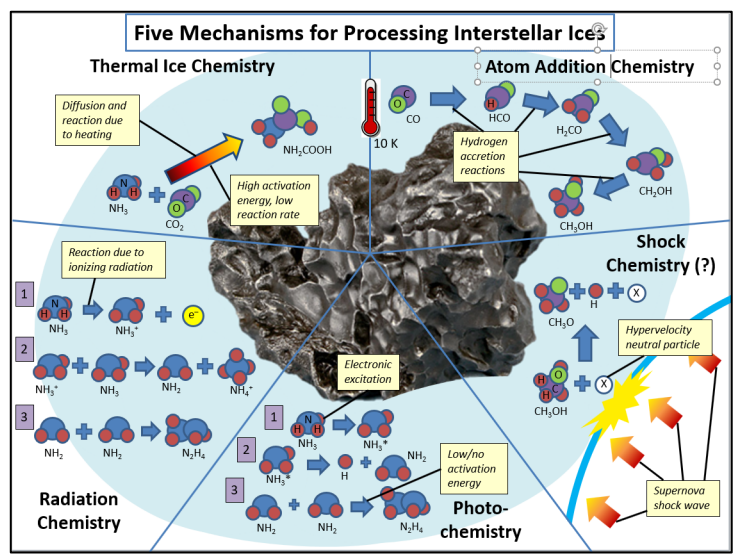

**Figure 5**
Non-energetic and energetic processing of interstellar ice.[3]

Photochemistry by UV light to produce radicals is one of five postulated mechanisms shown in Figure 5 for the processing of interstellar ices. These ~ 0.01 μm-thick ice mantles surrounding the micron-sized dust grains are found in pre-stellar cold cores and in the early stages of warm-up, leading to hot cores (e.g., Sagittarius B2N (Sgr B2N)) and hot corinos (e.g., L1551 IRS5).[i] Other mechanisms include purely thermal surface and bulk reactions, shock processes, radiolysis by highly energetic ions, and atom-addition chemistry.

The photochemistry induced by UV light produces both light and heavy radicals. While facile light radical diffusion is possible at ~10 K[27] (possibly leading to the atom addition chemistry shown in Figure 5), the gradual warm-up from ~10 K in cold cores to ~100 K in hot cores/hot corinos allows for heavy radical diffusion. The subsequent barrierless radical-radical reactions produce complex organic molecules (COMs), which likely lead to the synthesis of precursors of biologically relevant molecules.[j] The recent detection of COMs in starless and pre-stellar cold cores has also been attributed to photochemistry[28, 29] and radiation chemistry,[30] followed by non-diffusive mechanisms. Condensed phase photochemistry may also occur in protoplanetary disks, the planetary birthplaces that evolve from hot cores/corinos. Beyond the snowline in protoplanetary disks, in addition to the more dominant cosmic rays, the UV interstellar radiation field may process ices in which COMs such as methanol ($CH_3OH$), cyanoacetylene ($HC_3N$), and cyclopropenylidene ($c$-$C_3H_2$) have been discovered.

---

[i] Hot cores and hot corinos lead to high- and low-mass stars, respectively. The latter is often associated with planet formation.
[j] Some radical-radical reactions may be hindered by orientational effects.

**1.3.2. Condensed-phase photochemistry in interplanetary space**

Solar photons likely contribute to chemical synthesis in the icy bodies of the outer Solar System.[31] For example, while the detection of $H_2O_2$ on the surface of Europa has been attributed to water radiolysis initiated by energetic particles within Jupiter's magnetosphere, experimental results suggest that solar photons with energies below the threshold for ionization can penetrate meters below Europa's surface, facilitating the synthesis of $H_2O_2$.[32] Ice radiation chemistry and photochemistry may explain observations made by the *New Horizons* spacecraft during a close flyby of the Kuiper Belt object Pluto in 2015. Solar photons and resonantly-scattered interplanetary photons contribute to condensed-phase photochemistry, resulting in the synthesis of the observed red-colored macromolecular refractory tholins, the hydrolysis of which may lead to the production of amino acids and nucleobases on Pluto (Figure 6).[33]

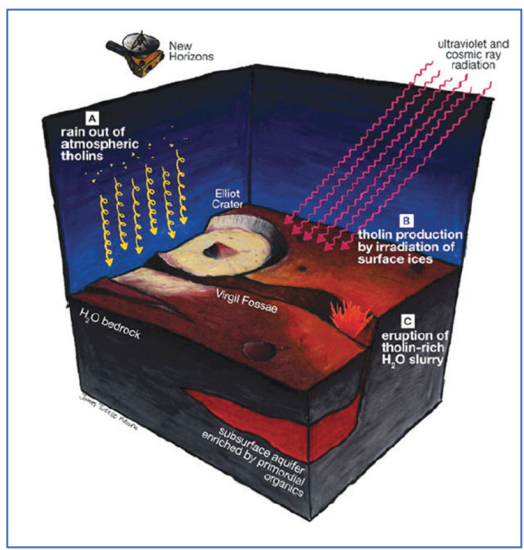

**Figure 6**
Tholin production by photochemistry of surface ices on Pluto.[33]
The publisher for this copyrighted material is Mary
Ann Liebert, Inc. publishers.

**2. Principles of Photochemistry**

**2.1. Photochemistry and electronically excited states**

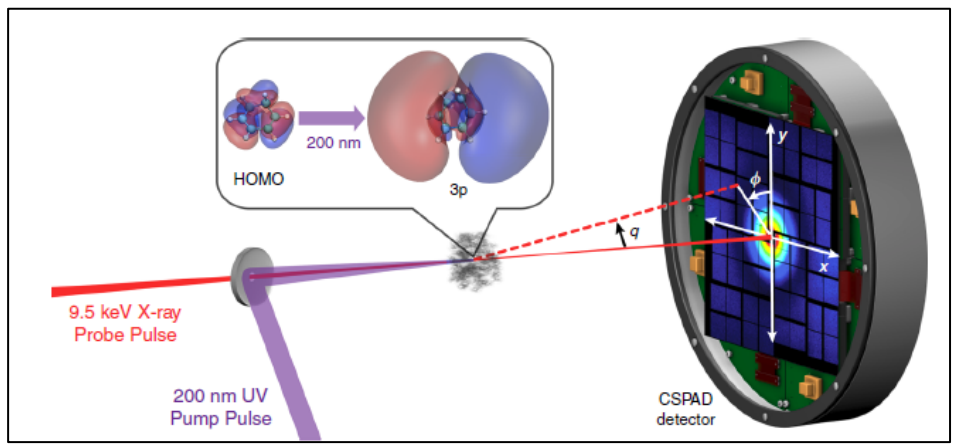

**Figure 7**
Change in the spatial distribution of electrons accompanying electronic excitation.[34]

Photochemistry is defined as "the branch of chemistry which relates to the interactions between matter and photons of visible[k] or ultraviolet[l] light and the subsequent physical and chemical processes which occur from the electronically excited state formed by photon absorption."[35] In contrast to radiation chemistry, photochemistry is associated with non-ionizing radiation.

The electronic rearrangement that accompanies the reactant molecule R being excited to R*, the first step in photochemistry ($R + h\nu \rightarrow R^*$), was recently captured in exquisite detail (Figure 7) using ultrafast x-ray scattering during which the movement of nuclei was negligible.[34] This change in the spatial distribution of electrons accompanying electronic excitation influences reactivity in profound ways. The second step involves the electronically excited molecule R* undergoing conversion to product(s), P.[m]

According to the IUPAC Gold Book, "Photochemical paths offer the advantage over thermal methods of forming thermodynamically disfavored products, overcome large activation barriers in a short period of time, and allow reactivity otherwise inaccessible by thermal methods." In contrast to thermal chemistry, which involves the ground electronic state, photochemistry concerns electronically excited states, each of which may have unique properties. The electronically excited molecule may decay via primary *photophysical*[n] and/or *photochemical* processes such as (1) luminescence (fluorescence or phosphorescence: ($A^* \rightarrow A + h\nu$)), (2) radiationless decay ($A^* + M \rightarrow A + M^*$), (3) quenching ($A^* + B \rightarrow A + B$), (4) photodissociation ($A^* \rightarrow B + C$), (5) photoisomerization ($A^* \rightarrow B$), (6) electron transfer ($A^* + B \rightarrow A^+ + B^-$), (7) excitation transfer or sensitization ($A^* + B \rightarrow A + B^*$), (8) bimolecular reaction ($A^* + B \rightarrow C + D$), (9) H-abstraction reaction ($A^* + RH \rightarrow AH + R$), and (10) ion-pair formation ($AB^* \rightarrow A^+ + B^-$).[o]

---

[k] 700–400 nm
[l] UV-A region (400–315 nm), UV-B (315–280 nm), and UV-C (280–100 nm)
[m] Excited states are stronger acids and stronger reductants than the original ground state.
[n] In a photophysical process, photon absorption leads to changes in electronic states without chemical transformation. Photophysical processes compete with photochemistry.

None of the above reactions of a photo-excited molecule involves secondary-electron production, a signature characteristic of radiation chemistry. Autoionization, which occurs only above the ionization threshold, makes a negligible contribution to the production of secondary electrons because it is a resonant process. Even if a local autoionizing resonance may have a much larger peak cross-section than the background normal-ionization continuum, the contribution from the ionization continuum will be much higher when the cross-sections are integrated over a wavelength region as short as 5 nm. Above the ionization threshold, direct ionization is much more likely than autoionization. Ion-pair production from excited electronic states also occurs only above ~ 10 eV and typically constitutes less than 0.1% of photoabsorption.[36] Therefore, reactions attributable solely to photochemistry are limited to those initiated by photons associated with far (deep)-UV (4.1–6.2 eV) (300–200 nm), near-UV (3.1–4.1 eV) (400–300 nm), and, occasionally, visible (1.8–3.1 eV) light.[p] Vacuum ultraviolet (VUV) (6.2–12.4 eV) light may initiate radiation chemistry, in addition to photochemistry. In contrast, in most cases, extreme ultraviolet light (EUV) (10.0–124 eV) triggers radiation chemistry, which involves ionization.[3]

## 2.2. Photochemistry vs. radiation chemistry

The energetic processing of interstellar ices to form prebiotic molecules may also occur via radiation chemistry, which is defined as the "study of the chemical changes produced by the absorption of radiation of sufficiently high energy to produce ionization."[37] Ionizing radiation in cosmic chemistry includes high-energy particles (e.g., cosmic rays) and high-energy photons (e.g., X-rays, which are thought to play a dominant role in the processing of protoplanetary disk ices close to the central star).[38] In addition to initiating chemical reactions in interstellar ices, ionizing radiation such as extreme ultraviolet (EUV) photons may affect the physical structure of interstellar ices via sputtering, lattice defect production, and non-thermal desorption. At energies above the ionization threshold (~ 10 eV), photon interactions with matter occur via three processes: (1) the photoelectric effect, (2) the Compton effect, or (3) pair production. In contrast to photochemistry, radiation chemistry is characterized by four phenomena: (1) production of a cascade of low-energy (< 20 eV) secondary electrons which are thought to be the dominant driving force behind radiation chemistry, (2) reactions initiated by cations, (3) non-uniform distribution of reaction intermediates,[q] and (4) non-selective chemistry leading to the production of multiple reaction products.[r3]

---

[o] We focus on photodissociation to produce radicals because current astrochemical simulations do not invoke photosensitization, abstraction reactions, photoaddition, or photocyclization.
[p] 1eV ≈ 100 kJ/mole ≈ 24 kcal/mole

The production of low-energy secondary electrons during radiation chemistry may lead to reaction pathways not available to photochemistry.[3] For example, while photon-induced singlet-to-triplet transitions are forbidden in the gas phase, electron-induced singlet-to-triplet transitions are allowed because the incident electron can be exchanged with those of the target molecule. Moreover, unlike photons, electrons can be captured into resonant negative ion states[s] that dissociate. The resulting molecular fragments may then react with the parent molecule or other daughter products to yield products unique to electron irradiation. Electron-induced reactions may predominate over photon-induced reactions because of the sheer number of low-energy secondary electrons produced by high-energy irradiation. In contrast to photon excitation, electron excitation is not a resonant process; the incident electron transfers the fraction of its energy sufficient to excite the molecule, and any excess is removed by the scattered electron.[3] This phenomenon enhances electron-induced reactions vis-à-vis photon-induced reactions. Reaction cross-sections can be several orders of magnitude larger for electrons than for photons, particularly at incident energies corresponding to resonances associated with dissociative electron attachment.[39]

### 2.3. Laws of photochemistry

Several core principles govern photochemistry. According to the Grotthuss-Draper law of photochemistry, only the light absorbed is effective in promoting photochemistry.[t] Most photochemical reactions obey the Bunsen-

---

[q] The interaction of ionizing radiation with matter produces tracks (events a few angstroms apart) and spurs (events hundreds of angstroms apart).
[r] In contrast, photochemistry can be exploited to trigger and control chemical reactions to yield specific products in the absence of side products.
[s] Electron attachment to form $AB^-$, a transient negative ion (TNI) occurs at low electron energies, typically below ~10 eV. The formation of a TNI is a resonant process because the final state ($AB^{-*}$) is a discrete state. Dissociative electron attachment (DEA) occurs when the TNI undergoes bond scission, resulting in an anion ($B^-$) in addition to a neutral atom/radical (A•).
[t] This law may appear self-evident but note that in stimulated emission, a molecule emits radiation that it had not previously absorbed.

Roscoe law, according to which a photochemical effect is directly proportional to the total energy dose, irrespective of the time required to deliver that dose.[u] In other words, photochemical effects, in contrast to some radiation chemistry outcomes,[v] depend on the photon fluence (photons per unit area, integrated over time) but not on the flux (photons per unit area per unit time). The Bunsen-Roscoe law is assumed valid for ice processing when extrapolating laboratory simulations (flux ~ $10^{14}$ photons/cm$^2$/sec) to dark, dense molecular clouds (flux ~ $10^4$ photons/cm$^2$/sec). According to the Stark-Einstein law, a single molecule absorbs only one photon in the primary step of a photochemical reaction. Multi-photon processes typically occur in laboratory settings involving lasers with high incident photon fluxes. In contrast, photochemistry of interstellar ices is likely to occur via single-photon events because secondary UV photon flux is low within dark, dense molecular clouds. While in photochemistry, a single < 10 eV photon typically initiates one chemical reaction, in radiation chemistry, a single photon/charged particle initiates a cascade of energy-loss events such as ionization, excitation, and nuclear displacement.[3]

## 2.4. Quantum yields

Quantum yield ($\Phi$), the number of molecules of reactant R converted per photon absorbed, is used to quantify the efficiency of photochemistry reactions.

$$\Phi = \frac{\text{number of molecules of R converted}}{\text{number of photons absorbed by R}}$$

Because some photochemical reactions involve a chain reaction mechanism, quantum yields can be as high as $10^6$ for reactions such as the chlorination of alkanes. In contrast, because photophysical processes such as radiative deactivation (e.g., fluorescence, phosphorescence) and non-radiative deactivation (e.g., quenching, intersystem crossing, and degradation to heat by internal conversion) compete with photochemistry,[w] often the quantum yield can be significantly less than one. In the condensed phase, an additional competing mechanism for photochemistry is photodesorption,[x] a non-thermal desorption mechanism partially responsible for the presence of gas-phase molecules inside dark, dense molecular clouds.[y]

---

[u] This law may not apply to high photon flux experiments involving multiphoton processes.
[v] The dose-rate effect in radiation chemistry is often associated with high linear energy transfer radiation such as heavy ions which comprise a portion of cosmic ray particles.
[w] Competing processes for a photoexcited molecule have time scales that range from tens of femtoseconds (e.g., internal conversion such as vibrational relaxation) to tens of seconds (e.g., phosphorescence).

## 3. Photochemistry Dynamics

### 3.1. Potential energy curves give static, dynamic, and spectroscopic information

Photochemistry is best understood by examining potential energy surfaces (PESs), which describe the potential energy of a molecular system as a function of the positions of the constituent atoms. According to the Franck-Condon principle, nuclei do not move appreciably during an electronic transition because photon absorption occurs approximately a hundred times faster than nuclear motion. Therefore, photon absorption is represented as a vertical transition in a potential energy diagram.

In the adiabatic Born-Oppenheimer approximation, when the nuclei move, the electrons adjust almost instantaneously, and the nuclei can be assumed to experience a time-averaged potential from the electrons. This decoupling of nuclear and electronic motion is the basis for potential energy surfaces (PESs), which are vital for understanding the structure of molecules. Additionally, the evolution of nuclear degrees of freedom in a single electronic eigenstate in accordance with the Born-Oppenheimer approximation provides insight into the thermal chemical reactivity involving a transition state (saddle point) in the ground state. A system in an electronically excited state, however, will likely encounter regions on the excited-state PES with non-negligible coupling between nuclear and electronic degrees of freedom.[40] The adiabatic approximation is violated if two or more levels become so close that nuclear motion promotes an electronic transition from one state to another. A prototypical example of non-adiabatic behavior occurs near conical intersections, defined as crossings between potential energy surfaces of the same spin multiplicity.[z] Conical intersections, so-called because of their double cone topography in two dimensions, play a central role in photochemistry and photophysics because excited states access ground-state products or reactants via this funnel. Conical intersections are important because they allow for *very fast, non-radiative* access to ground states from excited states. It is essential to note that the Condon approximation, which states that the electronic coupling between two states is independent of nuclear coordinates because electronic transitions occur much faster than nuclear motion, breaks down at conical intersections. At these seams, the coupling between electronic states depends *heavily* on nuclear coordinates.[41]

---

[x] Photodesorption has a typical quantum yield of $10^{-3}$ to $10^{-4}$ molecules per incident photon.
[y] A somewhat more efficient non-thermal desorption mechanism is known as reactive desorption, in which the exothermicity of a surface chemical reaction is converted to sufficient energy to cause desorption of the product.
[z] In the adiabatic BO approximation, potential energy curves of the same symmetry do not cross ("avoided crossing").

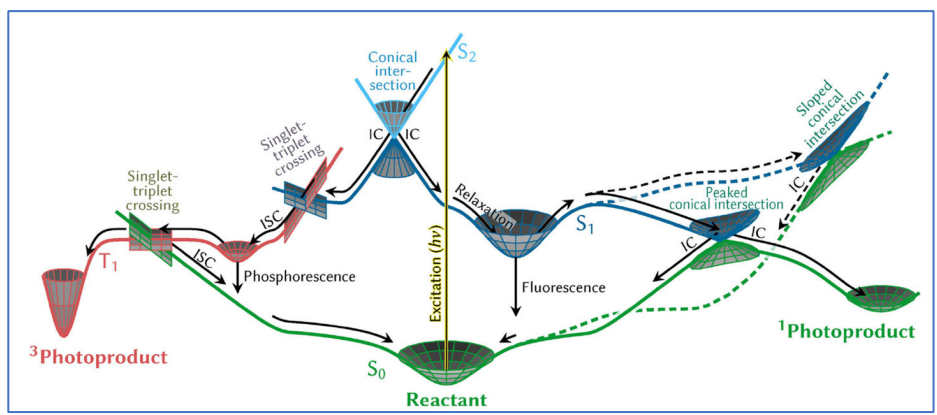

**Figure 8**
Temporal evolution of a photoexcited state.[42]

A difference in the equilibrium geometry in the ground and excited states results in the temporal evolution of the excited-state nuclear wave packet. As shown in Figure 8, the motion is dictated by the topography of the excited state potential energy surface.[42] In this specific example, following excitation from the ground singlet state ($S_0$) to the second excited singlet state ($S_2$), the system encounters the first excited singlet state ($S_1$) at a conical intersection. At this conical intersection, the system can either go right or left during internal conversion (IC). Along the right pathway, the system can either (1) relax to the $S_1$ minimum (right) and later undergo fluorescence or (2) overcome a barrier to reach either a peaked or a sloped conical intersection. From the peaked conical intersection, as shown in Figure 8, the system can evolve either toward the reactant or a singlet photoproduct. In contrast, from the sloped conical intersection, the nuclear wave packet reverts to the reactant. If the system evolves along the left branch after the first conical intersection, the nuclear wave packet undergoes a singlet-triplet transition (intersystem crossing (ISC)) to the $T_1$ state. Relaxation of the system leads to the $T_1$ state minimum from which phosphorescence may occur. If the wave packet escapes the minimum, it can evolve to either a triplet photoproduct or via an ISC revert back to the reactant (Figure 8).[42]

**3.2. Five scenarios for photon-induced dissociation of a diatomic molecule**

Depending on the shape of the excited state potential energy curves, photodissociation of a diatomic molecule may occur via (1) a bound upper state, (2) an unbound upper state, (3) predissociation, (4) indirect photodissociation, or (5) spontaneous radiative dissociation. Photodissociation is possible during an electronic transition to a state with bound energy levels if the transition is into the continuum part of the excited state spectrum (above the dissociation limit) (Figure 9).

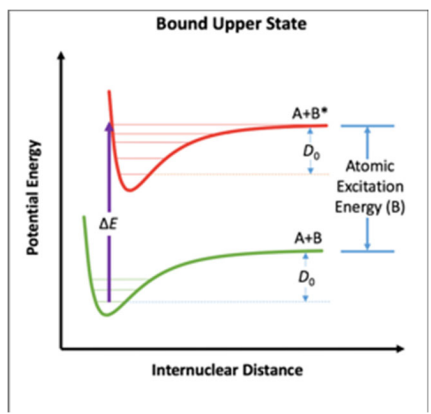

**Figure 9**
Photodissociation via a bound upper-state[3]

The other four photodissociation mechanisms are illustrated in Figure 10. In direct photodissociation (Figure 10 (A)), photoabsorption results in excitation to a repulsive state, leading to dissociation on the sub-picosecond timescale corresponding to a broad peak in the photodissociation cross-section as a function of wavelength.

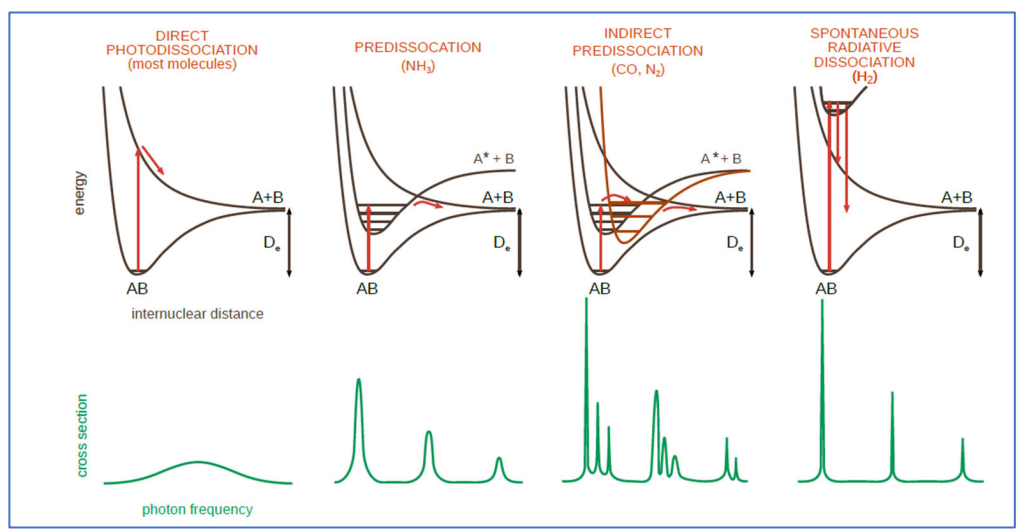

**Figure 10**
Other photodissociation mechanisms.[43] Reproduced with permission © ESO

If the potential energy curve of the excited electronic state is crossed by the curve of a repulsive excited state, direct predissociation (Figure 10 (B)) can occur. In the case of indirect predissociation (Figure 10 (C)), crossing to another bound state occurs before dissociation via a repulsive electronic state. In contrast to direct photodissociation, both types of predissociation result in photodissociation cross-sections with sharp peaks when plotted as a function of wavelength.[43] Predissociation occurring at different wavelengths for different isotopologues of the same molecule leads to isotope fractionation. Spontaneous radiative dissociation (Figure 10 (D)), the fifth possible mechanism for photodissociation (best exemplified by molecular hydrogen), involves excitation to a bound state followed by a radiative transition into an unbound electronic state, resulting in the decay of some fraction of the excited molecules by dissociation. This process allows $H_2$, the most abundant interstellar molecule, to shield itself on the edges of dense clouds against the ISRF.

**3.3. Selection rules**

Electronic excitation, the fundamental first step in photochemistry, is governed by spin and electric dipole transition selection rules, the latter of which are obtained from group theory symmetry arguments. The selection rules given below are best defined for gas-phase species. Strongly allowed transitions for a gaseous diatomic molecule depend upon the couplings among different angular momentum quantum numbers and whether the quantization is best treated in space-fixed or molecule-fixed coordinates. For the so-called pure Hund's case (a), transitions of linear molecules involve the following restrictions on angular momentum quantum numbers: (1) $\Delta \Lambda = 0, \pm 1$ ($\Lambda$ is the quantum number that specifies the total electronic orbital angular momentum along the molecular z-axis); (2) $\Delta S = 0$ ($S$ is the total spin angular momentum), and $\Delta \Omega = 0, \pm 1$ ($\Omega$ is the quantum number that specifies the total orbital angular momentum along the molecular z-axis). Electronic transitions in diatomic molecules are subject to additional restrictions resulting from their high degree of symmetry, as detailed elsewhere.[44,45] In molecular orbital parlance, $\sigma^* \leftarrow \sigma$ and $\pi^* \leftarrow \pi$ transitions are allowed but not $\pi^* \leftarrow \sigma$ or $\sigma^* \leftarrow \pi$. While the typical cross-section for allowed electronic transitions is $10^{-17}$–$10^{-18}$ cm$^2$, the nominally forbidden $\pi^* \leftarrow n$ transitions, involving excitation of non-bonding electrons to $\pi^*$ orbitals in species such as carbonyls, have significantly lower cross-sections. Solvent interactions cause $\pi^* \leftarrow \pi$ and $\pi^* \leftarrow n$ transitions to "redshift" and "blueshift," respectively.

In the gas phase, photon-induced processes are said to be spin-allowed when the spin multiplicity does not change (e.g., singlet to singlet and triplet to triplet transitions). For example, photon-induced gas-phase reactions of ammonia include spin-allowed (e.g., $NH_3 + h\nu \rightarrow NH(a^1\Delta) + H_2$) and disallowed (e.g., $NH_3 + h\nu \rightarrow NH(X^3\Sigma^-) + H_2$) processes. Because spin-orbit coupling is proportional to the fourth power of the atomic number, spin-forbidden transitions (e.g., singlet to triplet) become more likely for metal-containing heavy molecules in which the spin and orbital angular momenta are more strongly coupled. In contrast to the gas phase, spin-forbidden photon-induced transitions are allowed in the condensed phase.[46]

### 3.4. Application to gaseous molecular oxygen

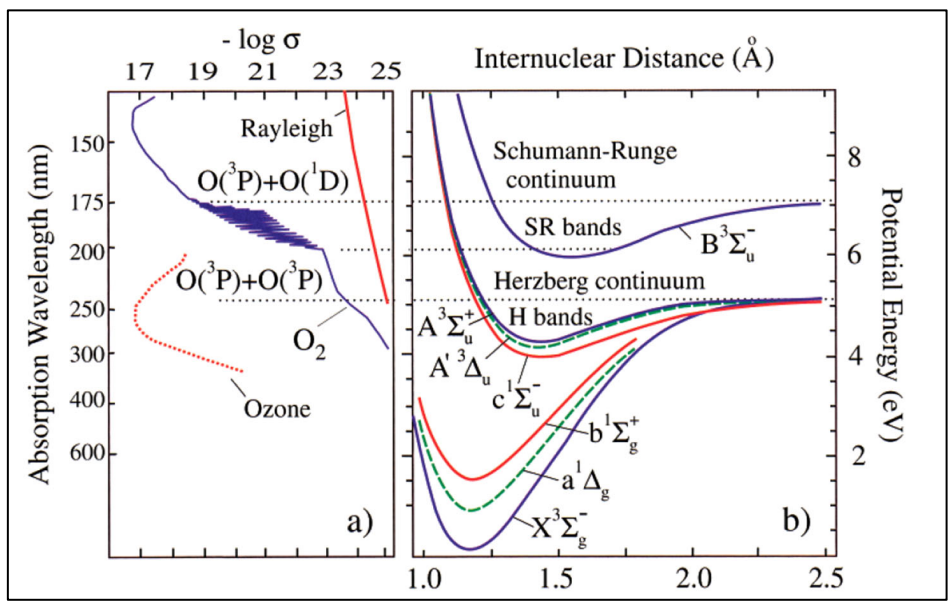

**Figure 11**
Bound potential-energy curves of molecular oxygen (right) and its UV absorption cross-section (left).[47]
Reprinted with permission. [47] Copyright 2000 American Chemical Society

Molecular oxygen is a prototypical example illustrating the dissociation mechanisms and the selection rules discussed above. Potential-energy curves for some $O_2$ bound states are plotted in Figure 11. According to the symmetry selection rules of optical transitions, absorption from the ground state, $X^3\Sigma_g^-$, into the next five lowest-energy electronic states is necessarily weak. Specifically, electric-dipole transitions between *gerade* states,[aa] required to excite $a^1\Delta_g$ and $b^1\Sigma_g^+$ excited electronic states from the ground state, are strictly forbidden, although weaker magnetic-dipole transitions generate an infrared spectrum.[48] Similarly, the $\Delta S=0$ selection rule forbids any absorption into the singlet $c^1\Sigma_u^-$ state from the triplet ground state, but the magnetic interaction of electron spins and orbital motion leads to weak

---

[aa] Parity of electronic state must change for a homonuclear diatomic molecule.

absorption. The electronic orbital angular momentum of the $A'^3\Delta_u$ states differs from the ground state by two quanta and the absorption into this electronic state is strongly forbidden, making the observed absorption extremely weak. Finally, single-photon absorption transitions are again forbidden between states of alike $\Sigma^+$ or $\Sigma^-$ symmetry, so the excitation to the $A^3\Sigma_u^+$ state is also excluded as a strong O$_2$ absorber.

The main absorption of O$_2$ then only occurs at wavelengths of 200 nm or shorter, corresponding to the photon energy needed to excite the $B^3\Sigma_g^-$ state, which has a set of quantum numbers fully compatible with absorption from the ground state.[47] Figure 11 shows the O$_2$ absorption cross-section at wavelengths matching the photon energies needed to excite molecular oxygen's electronic states. Direct dissociation into the $B^3\Sigma_g^-$ continuum leads to a slowly varying cross-section while absorption into bound levels between 200 and 175 nm produces a resonant structure. These are known respectively as the Schumann-Runge continuum and bands. Despite lying below the $B^3\Sigma_g^-$ dissociation energy, the Schumann-Runge bands are fully predissociated by additional repulsive states not shown in Figure 11. Forbidden absorption into the $A^3\Sigma_u^+$ state also leads to a "Herzberg" continuum between 250 and 200 nm. The Herzberg bands, a series of bands between 300 and 250 nm corresponding excitation to the $A^3\Sigma_u^+$ levels, lie below the ground state dissociation energy and are therefore strictly forbidden to dissociate.[47] The Herzberg continuum does not contribute to the UV opacity of the Earth's atmosphere because it is overlapped by far stronger ozone absorption, also plotted in Figure 11, while the Schumann-Runge continuum and bands dominate their part of the atmosphere's absorption spectrum, with the next-greatest contribution from Rayleigh scattering being negligible in comparison.[47]

**3.5. Photochemistry of polyatomic molecules**

| Dissociation channel | Notation in models | Branching ratio at 121.6 nm | Branching ratio at 118.2 nm |
|---|---|---|---|
| (1) $CH_3(X^2A_2'') + H$ | $CH_3$ | $\Phi(1) = 0.42 \pm 0.05$ | $\Phi(1) = 0.26 \pm 0.04$ |
| (2) $CH_2(a^1A_1) + H_2$ | $^1CH_2$ | $\Phi(2) = 0.48 \pm 0.05$ | $\Phi(2) + \Phi(3) = 0.17 \pm 0.05$ |
| (3) $CH_2(a^1A_1) + 2H$ | | $\Phi(3) \approx 0$ | |
| (4) $CH_2(b^1B_1) + H_2$ | – | $\Phi(4) \approx 0^a$ | $\Phi(4) \approx 0^a$ |
| (5) $CH_2(X^3B_1) + 2H$ | $^3CH_2$ | $\Phi(5) = 0.03 \pm 0.08$ | $\Phi(5) = 0.48 \pm 0.06$ |
| (6) $CH(X^2\Pi) + H + H_2$ | $CH$ | $\Phi(6) = 0.071^b$ | $\Phi(6) = 0.097^b$ |
| (7) $C(^1D) + 2H_2$ | – | $\Phi(7) = 0 \pm 0.006$ | $\Phi(7) = 0 \pm 0.006$ |

**Table 1**
Methane photochemistry branching ratios.[49]
Reprinted from[49] Copyright (2013), with permission from Elsevier.

Even for a relatively simple polyatomic molecule such as methane, the photochemistry of which plays a key role on Titan, VUV photodissociation involves complex phenomena. The allowed seven dissociation channels have branching ratios that are highly wavelength-dependent (Table 1) because of the Jahn-Teller distortion of the electronically-excited $^1T_2$ state of methane.[49] In systems with non-degenerate ground state vibronic levels but degenerate excited state vibronic levels such as $^1T_2$ of methane, stabilization is obtained through decreasing symmetry. Bond lengths within the molecule will distort to yield non-degenerate vibronic levels, causing band splitting in absorption spectra. In the case of methane, there are three distorted Jahn-Teller surfaces which produce three VUV absorption bands that likely make up the broad band in the VUV absorption spectrum of methane from 110–140 nm (8.9–11.3 eV).[49]

Methane photolysis may also involve the roaming mechanism, which is particularly important for formaldehyde, nitrate, and acetaldehyde gas-phase photochemistry. In a unimolecular reaction, the transition-state-theory (TST)-defying roaming mechanism involves the partial departure of an atom or molecular fragment from the molecule, forming a weakly bound radical-radical complex. The partially dissociated fragment possesses enough internal energy to roams up to several angstroms around the core in the van der Waals region.[50,51] At some point in the fragment's orbit, the fragment will abstract another atom, such as H (e.g., formaldehyde) or O (e.g., nitrate), from the core, forming a highly vibrationally-excited product. The recombination of the two radical fragments is nearly barrierless.[52] The roaming fragment bypasses the conventional saddle-point transition state and may take various trajectories with access to different transition states, but the potential energy surfaces over which it travels are always very shallow.[52,53] Based on recent studies, it has been claimed that any barrierless bond-cleavage process may be partly or mainly due to roaming.[51]

Despite its simplicity, methane poses an almost insurmountable challenge for providing a quantitative description of its photochemistry. The complexity arises from the need to follow the seven previously mentioned dissociation channels via adiabatic and non-adiabatic trajectories on a nine-dimensional potential energy surface.[49] For relatively complex molecules, a Jablonski diagram offers a simple schematic alternative to these trajectory calculations on multidimensional surfaces. In a Jablonski diagram, electronic and vibrational energy levels are shown as horizontal lines with no reference to relative nuclear coordinates. While radiative processes are depicted with straight arrows, non-radiative processes are illustrated with wavy arrows.

## 4. Photochemistry in the Condensed Phase

Photochemistry in the condensed phase may differ from that in the gas phase because the collective behavior of molecules in close proximity influences not only the absorption process but also the fate of the excited species. Figure 12 compares the absorption spectrum of molecular oxygen from 110–210 nm in the gas phase, in solid neon, and solid $O_2$ at 10 K. Because solvation reduces electronic state energies by different amounts, the absorption spectrum changes, causing spectra to be typically blue-shifted in the condensed phase. For condensed molecular oxygen, the primary absorption band is broadened and blue-shifted because the ground-state molecules interact more strongly with surrounding species in the solid phase than in the gas phase.[54] The solid $O_2$ absorption band, which peaks at ~7.0 eV is attributed to the absorption of the $(O_2)_2$ dimer.[54] This phenomenon is exclusive to the condensed phase. Absorption intensities may also be changed because interaction with neighboring molecules invalidates selection rules. In addition, quenching/collisional deactivation of the electronic excited state in the condensed phase is more likely than in the gas phase. Photochemistry yields in the condensed phase are diminished because of the so-called "cage effect," which allows two photofragments of a molecule to be trapped and undergo numerous collisions before they escape the cage. The "cage effect" may also hinder photon-induced geometrical rearrangement reactions.

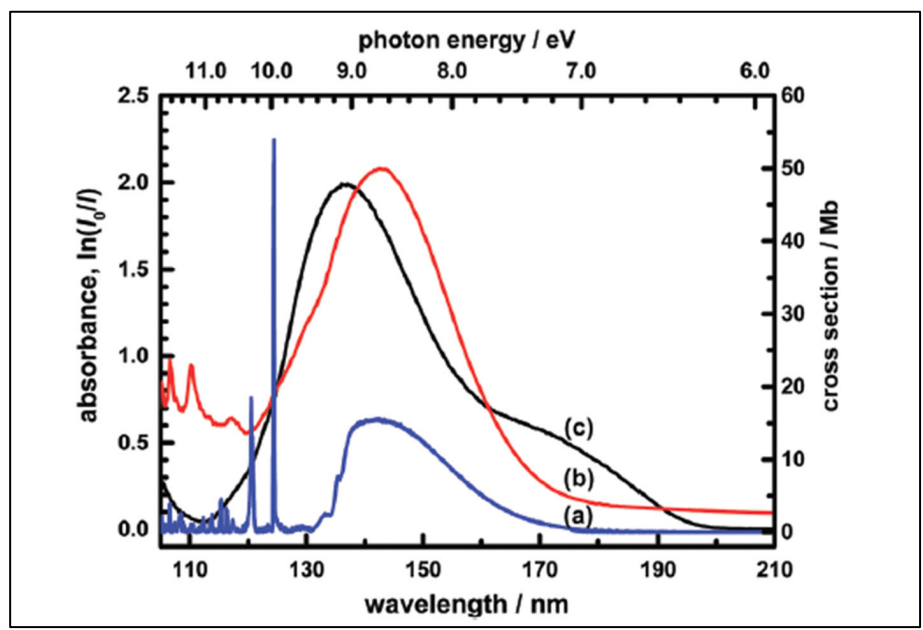

**Figure 12**
Absorption spectra of molecular oxygen (a) gas phase, (b) $O_2$ in solid Ne, (c) solid $O_2$ at 10 K.[55]

Alternatively, one can think about condensed-phase photochemistry using a diabatic basis, where each state can be directly assigned chemical significance (e.g., "reactants" live on State 1, and "products" live on State 2).[56] The kinetics and dynamics of condensed-phase photochemistry can be studied in a diabatic basis by applying one of various system-bath models, including the spin-boson model, the Redfield equation,[57] or the hierarchical equation of motion, among others.[58] In these models, the reactant and product states of interest for photochemistry are grouped together as the "system," and all solvent or phonon degrees of freedom are grouped together into a "bath." How the bath couples to the system is entirely defined by an object called the *spectral density*, which weights the different phonon or solvent vibrational modes in these condensed-phase systems by how much they directly influence the photochemical dynamics of the system of interest at a given temperature. The spectral density is related to the Fourier transform of the time autocorrelation function of a collective bath degree of freedom. Using these techniques, all of the effects that the condensed phase has on photochemistry, such as shortened excited-state lifetimes and blue-shifting, can be seen to arise due to either (1) the nonorthogonality of the diabatic basis or (2) vibrationally-mediated coupling between system and bath.[58]

**4.1. Exciton diffusion competing with photochemistry**

Condensed-phase photochemistry is influenced by the formation of excitons, which are localized electronically excited states associated with the generation of a negative electron and a positive hole (absence of an electron in the valence band) bound by Coulombic attraction. Excitons exist in nonmetallic crystals (insulators and semiconductors) and move freely as a unit, allowing for energy but not charge transportation. While exciton recombination results in photoemission, ice exciton diffusion from the bulk to the surface may cause desorption, lowering the efficiency of photochemical reactions in the condensed phase compared to the gas phase.[59]

**4.2. Excimers and exciplexes**

In contrast to gas-phase photochemistry, condensed-phase photochemistry may be altered by the formation of supramolecular excited collisional complexes whose lifetime is longer than those of the corresponding ground-state complexes. The excited supramolecular entity may either contain two chemically identical or different molecules, corresponding to excimers (**ex**cited d**imer**) or exciplexes (**ex**cited com**plexes**), respectively.[60] Pyrene $C_{16}H_{10}$, a polycyclic aromatic hydrocarbon (PAH)[bb] consisting of four fused benzene rings, provides an excellent example of these relatively long-lived molecular complexes of excited species.

**4.3. Kasha's rule**

In contrast to gas-phase photochemistry, during which molecular collisions are rare and energy dissipation is slow, condensed-phase photochemistry and photon emission should be independent of excitation energy because both processes occur mostly from the lowest excited state (Figure 13). This observation, known as Kasha's rule, can be rationalized by noting that vibrational relaxation (internal conversion) is a very rapid process in the condensed phase. While anti-Kasha gas-phase photochemistry is common, recent studies show examples of the breakdown of Kasha's rule in the condensed phase.[61] Kasha's rule is based on the intuition that internal conversion is always faster than any other process. If intersystem system crossing or electron transfer happens to be faster than internal conversion, then Kasha's rule will appear to be violated.

---

[bb] PAHs are thought to be the most ubiquitous class of large polyatomic molecules in space but no individual such species has yet been identified.

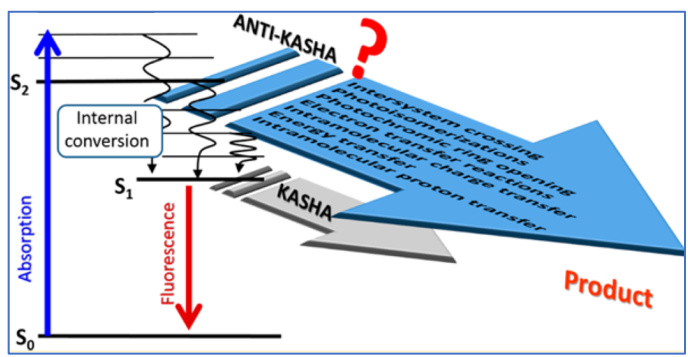

**Figure 13**
Anti-Kasha Photochemistry?[61]
Reprinted with permission.[61] Copyright 2017 American Chemical Society

### 4.4. Ionization energy in the condensed phase lowered because of screening

While ionization energy in the gaseous state is defined as the minimum energy required to remove an electron from a ground-state atom or ion, ionization energy in the condensed phase is defined as the energy difference between the vacuum level ($E_{vac}$) and the valence band maximum (VBM) as shown in Figure 14.[62] Vacuum-UV (6.2–12.4 eV) light may initiate radiation chemistry in addition to photochemistry because the threshold for producing secondary electrons is lower in the condensed phase than in the gas phase for a given molecule. For example, the photoelectric emission threshold for amorphous water ice (the main constituent of cosmic ices) is ~ 10.2 eV, which is smaller than the water gas-phase ionization energy of 12.6 eV. This reduction in threshold energy for ionization is a general phenomenon and has been ascribed to dielectric screening of the hole produced by UV irradiation in condensed matter.[62, 63]

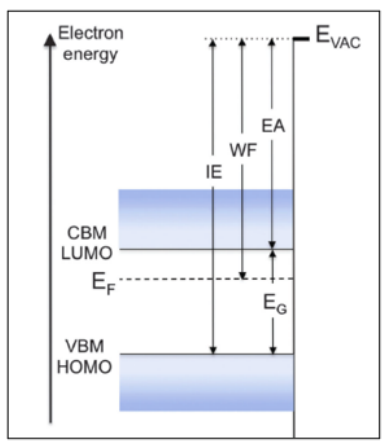

**Figure 14**

Definition of condensed phase ionization energy.[62] EF is the Fermi level, EG is the band gap, WF is the work function, and EA is the electronic affinity, IE is the ionization energy, and EVAC is the vacuum level.

### 4.5. Mean free path of photons in the condensed phase

In the condensed phase, 5–9 eV photons, those likely responsible for most solid-state photochemistry, have mean free paths ($\lambda$) comparable to the thickness of interstellar ice. For example, the mean free path of 8.5 eV photons in condensed water is ~ 0.06 microns, according to calculations based on the photon absorption cross-section ($5 \times 10^{-18}$ cm$^2$) of water ice. Because ice mantles surrounding cold interstellar grains are perhaps 1–100 monolayers, these ice mantles will be susceptible to photochemistry.

### 5. Conclusions

According to a recently published book, *The Stardust Revolution*, we are in the midst of the third scientific revolution, after those of Copernicus and Darwin.[64] This book espouses the astounding view that the origin of life can be traced back to the stars themselves. In 2016, unambiguous evidence for the presence of the amino acid glycine, an important prebiotic molecule, was deduced based on in situ mass-spectral studies of the coma surrounding cometary ice. This finding is significant because comets are thought to have preserved the icy grains originally found in the interstellar medium prior to solar system formation. Energetic processing of cosmic ices via photochemistry is thought to be a dominant mechanism for the extraterrestrial synthesis of prebiotic molecules. According to a 2019 PNAS publication, "Meteorites were carriers of prebiotic organic molecules to the early Earth; thus, the detection of extraterrestrial sugars in meteorites implies the possibility that extraterrestrial sugars may have contributed to forming functional biopolymers like RNA."[65]

**Acknowledgment**

Eric Herbst wishes to acknowledge the support of the National Science Foundation through grant AST19-06489 for his program in astrochemistry. EM gratefully acknowledges funding from the Arnold and Mabel Beckman Foundation. The Massachusetts Space Grant Consortium supported the work of MF. CRA's work was supported by grants from the National Science Foundation (NSF grant number CHE-1955215), Wellesley College (Faculty Awards and Brachman Hoffman small grants). AH work is supported by the CAAS ERDF/ESF "Centre of Advanced Applied Sciences" (No. CZ.02.1.01/0.0/0.0/16_019/0000778).